\documentclass[prl,twocolumn,aps]{revtex4}
\usepackage{graphicx}
\usepackage{bm}

\newcommand\ket{\rangle}
\newcommand\vk{\vec{k}}
\newcommand\up{\uparrow}
\newcommand\dn{\downarrow}
\newcommand\vS{\vec{S}}
\newcommand\vsig{\vec{\varsigma}}

\begin{document}

\title{Impurity Scattering Induced Entanglement of Ballistic
Electrons}
\author{A.T. Costa, Jr.}
\altaffiliation[Permanent address: ]{Departamento de Ci\^encias Exatas, Universidade
Federal de Lavras, 37200-000 Lavras, MG, Brazil.}
\email{a.costa@qubit.org}
\affiliation{Center for Quantum Computation, University of Oxford,
Clarendon Laboratory, Parks Road, OX1 3PU, Oxford, United Kingdom}
\author{S. Bose}
\email{sougato.bose@qubit.org}
\affiliation{Center for Quantum Computation, University of Oxford,
Clarendon Laboratory, Parks Road, OX1 3PU, Oxford, United Kingdom}

\begin{abstract}
We show how entanglement between two conduction electrons is
generated in the presence of a localized magnetic impurity
embedded in an otherwise ballistic conductor of special geometry.
This process is a generalization of beam-splitter mediated
entanglement generation schemes with a localized spin placed at
the site of the beam splitter. Our entangling scheme is
unconditional and robust to randomness of the initial state of the
impurity. The entangled state generated manifests itself in noise
reduction of spin-dependent currents.
\end{abstract}

\maketitle

In recent years, entanglement has emerged as a vital resource in
quantum information processing \cite{bennett00}. The controlled
generation of entanglement between constituents of a condensed
matter system would serve as a precursor to large scale quantum
computation in that system.  In this context, various methods of
generating entanglement between spin states of quantum dots
\cite{loss98,imamoglu99,hu01,mozyrski01}, electron spins
\cite{vrijen00}, electron numbers \cite{zanardi01},  nuclear
spins \cite{kane}, persistent current directions \cite{mooij99},
cooper-pair numbers \cite{makhlin99} and excitons
\cite{johnson99}, through controlled interactions between
relevant quantum systems have recently been proposed. Extraction
of entangled electrons from superconductors have been suggested
as well \cite{recher01}. Entanglement generation between
continuously interacting spins through the variation of
macroscopic parameters such as external fields and temperature
have also been studied \cite{Nat-ent}. Schemes also exist for
probing entangled states of electrons in solid state systems,
once such a state is available \cite{burkard00,loss00}.

  In this letter, we propose a scheme for
entangling the spins of two mobile electrons in a solid state
environment by scattering them off a localized magnetic impurity.
Being a scattering process, it has the advantage of not requiring
the careful switching on and off of interactions as in
Refs.\cite{loss98,imamoglu99,hu01,mozyrski01,vrijen00,kane,mooij99,makhlin99,johnson99}.
The interaction is automatically active {\em only when} the
electrons pass the impurity. Moreover, one may actually require
{\em mobile entangled electrons} for prototype demonstrations of
entanglement based quantum communication protocols
\cite{bennett00} or for detecting current noise based signatures
of entanglement \cite{burkard00}. Most known proposals would have
to first generate entangled electrons in an ``entangler" such as
a coupled dot system \cite{loss98,imamoglu99,hu01,mozyrski01}, a
spin resonance transistor \cite{vrijen00} or a superconductor
\cite{recher01} and then use additional processes to extract them
out into conducting leads as mobile electrons. Our scheme, on the
other hand, could {\em in principle}, be done entirely inside a
single ballistic conductor of special geometry. One suggestion for
entangling already mobile electrons does exist \cite{zanardi01},
but this does not entangle their spin degrees of freedom.
Compared to that proposal, ours has the advantage of extremely
long spin coherence times of conducting electrons
\cite{fabian98,tsukagoshi99}, and the potential to interface with
spin based solid state quantum computers
\cite{loss98,imamoglu99,hu01,mozyrski01,vrijen00,kane}.

A {\em second} motivation for our work is the generalization of
the existing beam-splitter mediated entanglement generation
schemes \cite{bose01}. In the scheme of Ref.\cite{bose01}, two
identical particles from uncorrelated sources can be made
entangled by using their indistinguishability and non absorbing
which-way detectors. One merely requires the particles to be
incident at a beam splitter simultaneously in oppositely spin
polarized (but disentangled) states. Our current proposal can be
regarded as a next step generalization of such a scheme with a
{\em localized spin placed at the site of the beam splitter}. The
consequences of this simple generalization, as we will show, are
quite profound. Firstly, the entanglement generation becomes {\em
deterministic} (though the degree depends on the coupling
strength of the localized spin with the incoming particles).
Secondly, {\em fermionic statistics} is used here in a {\em
fundamental} way to ensure that the incoming particles always
exit through different paths and the which-way detection of the
earlier scheme \cite{bose01} can be dispensed with. In contrast
to all other electron entangling methods
\cite{loss98,imamoglu99,hu01,mozyrski01,vrijen00,zanardi01,recher01},
our scheme, being based on a spin-spin scattering interaction and
fermionic statistics, should, with appropriate variations, be also
applicable to fermionic atoms and neutrons.

\begin{figure}
\includegraphics[scale=0.3]{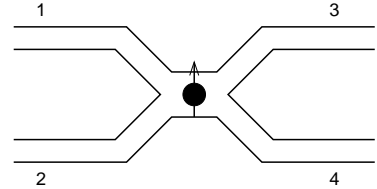}
\caption{Our electron entangling scheme. Two electrons in the same
spin state are injected into the rails $1$ and $2$. The electrons are
scattered by a magnetic impurity located at the meeting point of the
rails $1$ and $2$. As a result, the electrons always exit through separate
rails $3$ and $4$ in a spin entangled state.}
\label{scheme}
\end{figure}

Yet another {\em third} motivation for our work is predicting an
effect at the interface of two currently fashionable areas of
condensed matter physics. One of these areas is the manifestations
of fermionic statistics in two electron interference at beam
splitter-like mesoscopic structures \cite{yamamoto98}. The other
is the study of the {\em Kondo effect}, a phenomenon arising from
the interaction of the conduction electrons in a metal with a
localized magnetic impurity, in various mesoscopic systems
\cite{sasaki00,nygard00,madhavan98,odom00}. Our scheme simply
combines fermionic statistics at beam splitter-like structures
with Kondo-like scattering of conduction electrons from an
impurity (though, operating in a regime different to that of the
Kondo effect) to generate spin entangled electrons. Such a {\em
mobile} entangled state of electrons would then automatically
manifest itself through noise reduction of spin-dependent
currents in appropriate conductor geometries, as pointed out by
Burkard {\em et. al.} \cite{burkard00}.

   Our setup consists of a ballistic conductor with four rails $1,2,3$ and
$4$ meeting at a common junction as shown in Fig.\ref{scheme}. In
order to generate entanglement, two conduction electrons are
injected in the rails $1$ and $2$. Both the electrons must have
the same spin orientation, which can be achieved by the use of a
spin filter as part of the injection mechanism. The spot where the
rails $1$ and $2$ meet has a localized impurity atom as shown in
Fig.\ref{scheme}. In the regime where the impurity atom has a net
magnetic moment, the conduction electrons will be magnetically
scattered by the impurity. As a result, the two output electrons
{\em always exit} the scattering region through {\em separate}
rails $3$ and $4$ and are {\em always entangled}. The degree of
entanglement depends on the strength of the scattering
interaction. Moreover, if the spin of the magnetic impurity can
be measured, then, conditional on a spin-flip of the impurity, the
electrons in exit rails $3$ and $4$ are maximally entangled.


We are going to consider a simplified model of the proposed
device. We assume the rails to be metallic with the electronic
states being treated as those of a one dimensional gas of
non-interacting electrons, and the impurity will be represented
by a localized orbital. A host metal containing an impurity atom
with an incomplete $d$ or $f$ shell can be described by the well
known Anderson hamiltonian \cite{hewson}. For certain values of
the parameters, the formation of a localized magnetic moment at
the site of the impurity is favored. In this case the Anderson
hamiltonian can, by means of the Schrieffer-Wolff transformation
\cite{hewson}, be cast in to the form
\begin{eqnarray}
H_{s-d}=\sum_{\vk,\sigma}\varepsilon(\vk)a^\dagger_{\vk\sigma}a_{\vk\sigma} +
J\sum_{\vk,\vk^\prime}\vS . \vsig_{\vk,\vk^\prime}\, ,\nonumber\\
\vsig_{\vk,\vk^\prime} =
{\rm a}^\dagger_{\vk}\vec{\sigma}{\rm a}_{\vk^\prime}\, ,
\label{kondo}
\end{eqnarray}
which is the so called $s-d$, or Kondo, hamiltonian. Here,
$\varepsilon(\vk)$ is the dispersion relation for the conduction
electron gas, $a^\dagger_{\vk\sigma}$ creates a conduction
electron with spin $\sigma$ and wave vector $\vk$,
$\vS$ is the spin operator of the impurity, $\vec{\sigma}$ is the vector
formed by the Pauli matrices, ${\rm
a}_{\vk}^\dagger=(a^\dagger_{\vk\up},a^\dagger_{\vk\dn})$, and $J$
is the exchange coupling per atom between the localized spin and
the conduction electrons' spin \cite{exchange}.
Defining the spin raising and lowering operators in the usual way:
$\varsigma^+ = a^\dagger_\up a_\dn$, $\varsigma^-=a^\dagger_\dn a_\up$,
$S^+ = S^x + iS^y$, $S^- =S^x - iS^y$, one can write the interaction term
${\cal V}$ as a sum of a ``longitudinal'' and a ``transverse'' part, ${\cal V} = {\cal
V}_{||} + {\cal V}_{\perp}$. The longitudinal part ${\cal
V}_\perp= S^+\varsigma^- + S^-\varsigma^+$ is responsible for
spin-flip processes, in which the spin of a conduction electron
and of the localized state are both flipped.

We are interested in what happens to a conduction electron
propagating through the system depicted in Fig. \ref{scheme}. We
will assume that the rails are of reduced transverse dimensions,
such that there are only very few conducting states near the
Fermi level. This allows us to simplify the analysis of the
conduction process by assuming that only one channel is important
in their description. If we disregard, for the moment, the
presence of the magnetic impurity, the geometry of the rails is
that of a ``beam splitter.'' Following Loudon
\cite{loudon,buttiker2}, the propagation of the conduction
electrons through this kind of system can be described by a
scattering matrix ${\bf s}$, which relates the ``incoming''
states (electrons propagating on rails 1 and 2) to the
``outgoing'' states (electrons propagating on rails 3 and 4),
$\alpha_l=\sum_{m=3,4}s_{lm}\alpha_m$, where
$\alpha^\dagger_{l\sigma}$ creates an electron with spin $\sigma$
in the propagating channel of rail $l$. In principle, the only
requirement on ${\bf s}$ is unitarity, and the explicit form of
its elements is determined by the transmission properties of the
system.

The role of the magnetic impurity will be described by the $s-d$ hamiltonian
\ref{kondo} taking into account the specific characteristics of the proposed
geometry. In the end we will rewrite the final state in terms of the operators
$\alpha^\dagger_{l\sigma}$, which create conduction electrons with
spin $\sigma$ propagating in rail $l$ with Fermi wave vector $\vk_F$.
We shall assume that all the other electronic states in the rails are
occupied, and play no role in the propagation process. Thus, the
scattering of the conduction electrons by
the impurity may be described by the $T$-matrix associated with
$\cal V$ in the first Born approximation \cite{yosida},
\begin{eqnarray}
T^{(1)} = {\cal V} =J\sum_{l=1,2}\sum_{\vk,\vk^\prime}
\{ S^+a^\dagger_{l\vk\dn}a_{l\vk^\prime\up} +
   S^-a^\dagger_{l\vk\up}a_{l\vk^\prime\dn} + \nonumber\\
   S^z\lbrack a^\dagger_{l\vk\up}a_{l\vk^\prime\up}-
              a^\dagger_{l\vk\dn}a_{l\vk^\prime\dn}\rbrack\}\, .
\label{tmatrix}
\end{eqnarray}

This $T$-matrix may now be used to calculate the final scattering
state. In momentum space \cite{yosida}, $|\vk\ket^+ =
\sum_{\vk^\prime}{\cal S}_{\vk^\prime\vk}|\vk^\prime\ket$, and
${\cal S}_{\vk^\prime\vk} = \delta_{\vk^\prime\vk} - 2\pi
i\delta(\epsilon(\vk^\prime)-\epsilon(\vk))T_{\vk^\prime\vk}$. A
straightforward calculation shows that, if one takes as the
initial state
$|\vk_F\ket=a^\dagger_{1\vk_F\up}a^\dagger_{2\vk_F\up}|0\ket\otimes|\dn\ket$,
with  $|\dn\ket=d^\dagger_\dn|0\ket$, the (unnormalized) final
state,
\begin{eqnarray}
|\vk_F\ket^+ = (1+i\bm{J})|\up\up\ket\otimes|\dn\ket
-2\sqrt{2}i\bm{J}|\psi^+\ket\otimes|\up\ket\, .
\end{eqnarray}
where $\alpha_{l\sigma}=a_{l\vk_F\sigma}$, $\bm{J}=\pi J\rho(\epsilon_F)$,
$|\up\up\ket = \alpha^\dagger_{3\up}\alpha^\dagger_{4\up}|0\ket$,
$|\psi^+\ket=\frac{1}{\sqrt{2}}(\alpha^\dagger_{3\up}\alpha^\dagger_{4\dn} +
\alpha^\dagger_{3\dn}\alpha^\dagger_{4\up})|0\ket$
and $\rho(\epsilon_F)$ is the density of states for the
conduction electrons at the Fermi level in the rails.
It should be noted that we are assuming that the temperature is
larger than Kondo temperature, in which case we do not need to worry
about the breakdown of the perturbation expansion of the $T$ matrix.

If one considers higher order terms in the $T$ matrix expansion the result is not
qualitatively different. It can be easily shown that the contributions from the
higher orders factor out, and the final state still has a component corresponding
to the initial state plus a maximally entangled component with a weight of the
order of $J\rho(\epsilon_F)$.

The value of $J\rho(\epsilon_F)$ can be estimated as follows \cite{exchange}:
bearing in mind that all the energies involved are of the order of the width
of the conduction band $D$, one readily verifies that $J\sim D$.
$\rho(\epsilon_F)$ can be roughly estimated for a normal metal by considering a
constant density of states extending over the band width $D$,
which gives $\rho(\epsilon_F)\sim \frac{1}{D}$. Thus, $\bm{J}=\pi
J\rho(\epsilon_F)\sim 1$, and the weight of the entangled part of
the final state is of the same order as that of the non-entangled
component.

\begin{figure}
\includegraphics[scale=0.3]{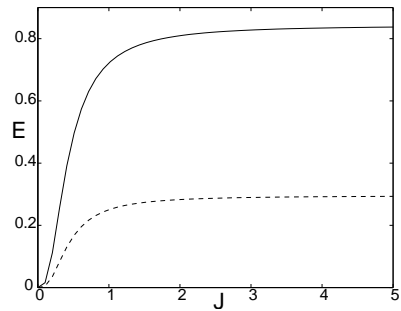}
\caption{The entanglement of the electrons in the output rails
$3$ and $4$ as a function of $\bm{J}$. The solid line is the entanglement
for a definite initial state of the impurity spin. The dashed line shows
the entanglement for a random initial state of the impurity spin. }
\label{fig2}
\end{figure}

The method of obtaining the highest amount of entanglement would
now be to measure the spin of the impurity atom after the
electrons have scattered. If the spin of the impurity is measured
and found flipped, the electrons in the rails $3$ and $4$ will be
projected onto the maximally entangled state $|\psi^{+}\rangle$.
The probability for this to happen is $8\bm{J}^2/(1+9 \bm{J}^2)$,
which is finite for any non-zero $\bm{J}$. However, as far as the
generation of entanglement is concerned, our scheme is {\em
unconditional}. Even if the impurity spin was {\em not measured
at all}, the electrons are projected onto the mixed state
\begin{equation}
\Lambda= \frac{1+\bm{J}^2}{1+9 \bm{J}^2}|\uparrow\uparrow\rangle
\langle \uparrow\uparrow| + \frac{8
\bm{J}^2}{1+9\bm{J}^2}|\psi^{+}\rangle\langle\psi^{+}|,
\end{equation}
which is entangled {\em irrespective} of the value of $\bm{J}$.
The entanglement of the above state for a range of feasible (of
the order of unity) values of $\bm{J}$, can be calculated from a
formula by Wootters \cite{wootters98} and is shown in
Fig.\ref{fig2}. The plot clearly shows that entanglement is
already above $0.8$ for $\bm{J}\sim 3$. Apart from being
unconditional, our scheme also shows {\em robustness} to
uncertainty in the initial spin direction of the impurity. In
fact, if the impurity spin is initially in the completely random
state $|\uparrow\rangle \langle \uparrow| +|\downarrow\rangle
\langle \downarrow|$, the final state $\Lambda^{'}=
(1+5\bm{J}^2)/(1+9\bm{J}^2)|\uparrow\uparrow\rangle \langle
\uparrow\uparrow| + 4\bm{J}^2/(1+9 \bm{J}^2)|\psi^{+}
\rangle\langle\psi^{+}|$, of the electrons is {\em still
entangled}. This is also plotted in Fig.\ref{fig2} as a dashed
line.

  For detection of the above entangled mixed state $\Lambda$ (or $\Lambda^{'}$), one can use a simple
modification of the current noise based method suggested by
Burkard {\em et. al.} \cite{burkard00}. As in
Ref.\cite{burkard00}, the electrons in rails $3$ and $4$ should
be brought together to interfere at a beam splitter. The currents
in the outputs of the beam splitter (say, rails $5$ and $6$) will
be {\em completely noiseless} for the state $\Lambda$ we have
produced in the rails $3$ and $4$. This fact, in itself, is {\em
not sufficient} to guarantee that the state $\Lambda$ is
entangled. One then needs to measure the {\em spin correlation}
$\langle S_z(5) S_z(6) \rangle$ of the electrons coming through
the rails $5$ and $6$. No disentangled mixed state (a state of the
form $\sum_i p_i |\psi_i\rangle \langle
\psi_i|\otimes|\phi_i\rangle \langle \phi_i|$) in the rails $3$
and $4$ can produce zero current noise in the outputs $5$ and $6$,
{\em unless} $|\psi_i\rangle=|\phi_i\rangle$ for all $i$. This
means that noiseless current in rails $5$ and $6$ is consistent
with a disentangled state in rails $3$ and $4$ {\em only when}
$\langle S_z(5) S_z(6) \rangle=1$. However, for our state
$\Lambda$, we will measure $\langle S_z(5) S_z(6) \rangle$ to be
$(1-7\bm{J}^2)/(1+9 \bm{J}^2)$, which {\em guarantees} its
entanglement for any nonzero $\bm{J}$.


We now discuss the issue of feasibility. We can choose the rails for
the electrons to
be carbon nanotubes, which are ballistic \cite{todorov98,liang01}
and spin coherent \cite{tsukagoshi99} conductors. Cross junctions
of carbon nanotubes, as required by us, have been fabricated
\cite{fuhrer00,yjunct}. In the context of Kondo effect experiments in
nanotubes, recently cobalt clusters have been embedded in
nanotubes as magnetic impurities \cite{odom00}. If one can
similarly place a single cobalt atom, it would serve the purpose
of the localized spin in our experiment.
 In general, the
effect of impurities in carbon nanotubes are averaged over the
circumference of the entire tube \cite{todorov98} and sudden
narrowing of the tube at the site of the localized spin could
increase the scattering strength (our $J$). Alternatively, a
junction between a nanotube and a very small conductor of a real
metal can be made (such as the electrode-nanotube junctions in
Ref.\cite{liang01}), and the impurity can be placed in this small
length of metal. Another option comes from the recent
implementation of Kondo effect in quantum dot carbon nanotubes
\cite{nygard00}. Conducting nanotubes can be connected to a
quantum dot nanotube, whose spin effectively serves as our
magnetic impurity. Quite apart from carbon nanotubes, one can
have an all semiconductor implementation of our proposal by
modifying the setup of Ref.\cite{yamamoto98} by placing a quantum
dot with spin at the site of the beam-splitter. Indeed, such
semiconductor quantum dots have served as localized magnetic spins
in recent Kondo experiments \cite{sasaki00}. Moreover, an all
metal implementation is also possible if small enough gold wires
can be fabricated so that electron transport in them is
ballistic. This will then have to be combined with the deposition
of a single cobalt atom on a gold substrate (as in
Ref.\cite{madhavan98}) to obtain our setup. Quite outside the
realm of electrons, ballistic waveguides and beam splitter like
microstructures have been recently fabricated for tests of atomic
statistics \cite{schmiedmayer00}. If one can localize an atomic
spin at the beam splitter, then fermionic atoms could also be
used to carry out our experiment.

  In this letter, we have presented a scheme for entangling the
spins of two conducting electrons which uses the combined effects
of magnetic scattering and fermionic statistics. It has
advantages of obtaining the entangled electrons mobile in separate
wires, not requiring control over spin-spin interactions, not
requiring non-absorbing measurements of electron paths, and is
applicable to all fermions. The entangling is successful
irrespective of the final state of the impurity and is robust to
uncertainty about the initial state of the impurity. Further work
can focus on the possibility of using electron scattering from
successive magnetic impurities to implement a two qubit logic
gate between the impurity spins. Prototype implementations of
quantum communications through magnetic scattering and entangled
electrons could also be studied. In particular, some quantum
information processing protocols using statistics in a
fundamental way have been proposed recently \cite{omar01}.
Extensions of these schemes with an additional magnetic
scattering should be interesting.

\acknowledgments{
A.T.C. gratefully acknowledges financial support from CNPq - Brazil.
}



\begin{thebibliography}{99}
\bibitem{exchange} As a result of the Schrieffer-Wolf transformation,
it can be shown that $J=\frac{V^2U}{(U+\epsilon_d-\epsilon_F)(\epsilon_F-\epsilon_d)}$,
where $U$ is the effective Coulomb interaction, $V$ is the $s-d$ mixing parameter, $\epsilon_F$
is the Fermi energy of conduction electrons and $\epsilon_d$ is the on-site energy of
the $d$ level. The Kondo regime is characterized by values of those parameters such that
$J>0$.
\end{thebibliography}
\end{document}